\documentclass[conference]{IEEEtran}
\IEEEoverridecommandlockouts

\usepackage{cite}
\usepackage[hidelinks]{hyperref}
\usepackage{amsmath,amssymb,amsfonts}
\usepackage{multirow}
\usepackage{multicol}
\usepackage{makecell}
\usepackage{pifont}
\usepackage{stfloats}
\usepackage{longtable}

\usepackage[most]{tcolorbox}

\usepackage{algorithmic}
\usepackage{graphicx}
\usepackage{textcomp}
\usepackage{xcolor}
\def\BibTeX{{\rm B\kern-.05em{\sc i\kern-.025em b}\kern-.08em
    T\kern-.1667em\lower.7ex\hbox{E}\kern-.125emX}}

\makeatletter
\newcommand{\linebreakand}{%
  \end{@IEEEauthorhalign}
  \hfill\mbox{}\par
  \mbox{}\hfill\begin{@IEEEauthorhalign}
}
\makeatother

\begin{document}

\title{Enhancing Large Language Models \\for Automated Homework Assessment \\in Undergraduate Circuit Analysis}

\author{
\IEEEauthorblockN{Liangliang Chen, Huiru Xie, Zhihao Qin, Yiming Guo, Jacqueline Rohde, Ying Zhang$^{\dagger}$\thanks{\textcopyright 2025 IEEE. Accepted to 2025 Frontiers in Education (FIE) Conference. Personal use of this material is permitted. Permission from IEEE must be obtained for all other uses, in any current or future media, including reprinting/republishing this material for advertising or promotional purposes, creating new collective works, for resale or redistribution to servers or lists, or reuse of any copyrighted component of this work in other works.}}
\IEEEauthorblockA{\textit{School of Electrical and Computer Engineering}, \textit{Georgia Institute of Technology}, Atlanta, USA \\
$^{\dagger}$Corresponding Author. E-mail: \texttt{yzhang@gatech.edu}}}

\maketitle

\begin{abstract}
This research full paper presents an enhancement pipeline for large language models (LLMs) in assessing homework for an undergraduate circuit analysis course, aiming to improve LLMs' capacity to provide personalized support to electrical engineering students. Existing evaluations have demonstrated that GPT-4o possesses promising capabilities in assessing student homework in this domain. Building on these findings, we enhance GPT-4o’s performance through multi-step prompting, contextual data augmentation, and the incorporation of targeted hints. These strategies effectively address common errors observed in GPT-4o's responses when using simple prompts, leading to a substantial improvement in assessment accuracy. Specifically, the correct response rate for GPT-4o increases from 74.71\% to 97.70\% after applying the enhanced prompting and augmented data on entry-level circuit analysis topics. This work lays a foundation for the effective integration of LLMs into circuit analysis instruction and, more broadly, into engineering education.
\end{abstract}


\section{Introduction}

Large language models (LLMs) are rapidly transforming a wide range of fields, including programming \cite{xu2022systematic}, robotics \cite{ma2023eureka, chen2024rlingua}, and education \cite{kasneci2023chatgpt}. Pre-trained on vast corpora spanning diverse domains, LLMs demonstrate impressive question-answering capabilities across numerous benchmarks \cite{achiam2023gpt, touvron2023llama, chen2025benchmarking}. Furthermore, their use of natural language inputs and outputs makes them highly accessible, particularly for users without specialized expertise. In the field of education, students' diverse backgrounds and learning preferences \cite{tetzlaff2021developing} create a demand for personalized instruction. This, in turn, places a heavy workload on instructors striving to provide effective learning support \cite{ujir2020teaching}. LLMs are well-suited to address this challenge, as they can deliver instant, personalized feedback, explanations, and tutoring at scale. Motivated by this potential, this paper investigates the application of LLMs in engineering education, with a specific focus on automated homework assessment in undergraduate circuit analysis.

Prior to the era of LLMs, the development of automated homework assessment tools had been a long-standing goal pursued by researchers for several decades. These tools primarily targeted disciplines or subjects where rule-based assessments could be readily implemented, such as programming \cite{ala2005survey}, spreadsheets \cite{blayney2004automated}, or tasks in natural language processing, such as automated writing evaluation \cite{warschauer2008automated}. The advent of LLMs has made automated homework assessment more flexible and less dependent on rigid rule-based systems, allowing their application to a broader range of disciplines \cite{chu2025llm}. In fact, the potential of LLMs has been explored across various branches of education. For example, Yan \textit{et al.} \cite{yan2025mathagent} developed a mixture-of-math-agent framework for multimodal mathematical error detection. 
Ref. \cite{tsai2023exploring} employed LLMs to foster hands-on problem-solving and programming skills in chemical engineering education, aiming to promote critical thinking and facilitate students' deeper understanding of core subjects. In the domain of analog circuits, Skelic \textit{et al.} \cite{skelic2025circuit} presented an LLM benchmark for circuit interpretation and reasoning. Using their constructed dataset, GPT-4o—identified as the best-performing model—achieved an accuracy of approximately 48\% when evaluated on final numerical answers, indicating that it remains insufficiently reliable for practical use in educational settings. Our work \cite{chen2025benchmarking} investigated the application of LLMs in automated homework assessment for undergraduate circuit analysis. Three models—GPT-3.5 Turbo, GPT-4o, and Llama 3 70B—were evaluated using a dataset of reference solutions and real student submissions covering key circuit analysis topics. Due to current limitations of LLMs in interpreting handwritten or printed images, the evaluations in \cite{chen2025benchmarking} relied on reference solutions as the ground truth. Five aspects of student work were assessed with a unified prompt: completeness, method, final answer, arithmetic accuracy, and unit correctness. Results show that GPT-4o and Llama 3 70B outperformed GPT-3.5 Turbo across all metrics, with each model exhibiting unique strengths. However, both GPT-4o and Llama 3 70B displayed recurring hallucinations, such as failures to recognize equivalent numerical formats, rounding errors, and alternative solution methods. Note that reliability is a critical consideration for educational support tools as incorrect responses from LLMs may mislead students. To the best of our knowledge, no prior work has focused on generating reliable homework assessments for undergraduate circuit analysis.

Building an LLM capable of reliably assessing students’ homework in circuit analysis presents several challenges:
\begin{itemize} 
\item[i)] Circuit analysis covers a broad range of concepts, including fundamental circuit elements, circuit laws and theorems, and various analytical techniques \cite{svoboda2013introduction}. To produce reliable responses, LLMs must not only leverage this domain-specific knowledge—presumed to be within the capabilities of advanced models such as GPT-4o—but also exhibit strong mathematical reasoning to perform accurate analyses. However, the limited reasoning capabilities of current LLMs \cite{frieder2023mathematical, sessler2024benchmarking} make this task particularly challenging, necessitating strategies either to address or circumvent these limitations.

\item[ii)] Circuit analysis homework often involves circuit diagrams. However, current state-of-the-art LLMs are not yet capable of reliably extracting the necessary information from these diagrams \cite{chen2025benchmarking}. As a result, LLMs need to rely solely on textual information to perform assessments. Thus, prompt design must ensure that all essential information from the diagrams is explicitly provided in text, preventing the model from making speculative inferences.

\item[iii)] Students may make errors across various dimensions, such as methodological flaws, arithmetic mistakes, or omissions of units. When evaluating the solutions, LLMs are prone to losing focus, especially in more complex or advanced topics, which can lead to inaccurate feedback.
\end{itemize}

The contributions of this paper are summarized as follows:

\begin{itemize} 
\item[i)] Based on the evaluation results in \cite{chen2025benchmarking}, GPT-4o is selected as the candidate LLM for developing a reliable automatic homework assessment tool due to its strong overall performance. To mitigate hallucinations, we adopt a multi-step prompting approach, where each step targets a specific aspect of the homework assessment. The responses generated for each aspect are subsequently integrated to produce a comprehensive overall evaluation. 
\item [ii)] Ref. \cite{chen2025benchmarking} summarizes the strengths and weaknesses of GPT-4o in generating assessments from various perspectives during LLM evaluations. In this paper, we address the weaknesses identified in \cite{chen2025benchmarking} by designing targeted hints and integrating them into specific steps of the LLM prompts to improve GPT-4o’s performance in its weaker areas. Additionally, we construct offline datasets to provide supplementary context for the tasks, thereby enhancing the reliability of GPT-4o's responses. Although the targeted hints are tailored to GPT-4o, the overall enhancement framework proposed in this paper is broadly applicable for addressing limitations in other LLMs and supporting similar improvements.
\end{itemize}

\begin{figure*}[!h]
\centering
\includegraphics[width=0.85\textwidth]{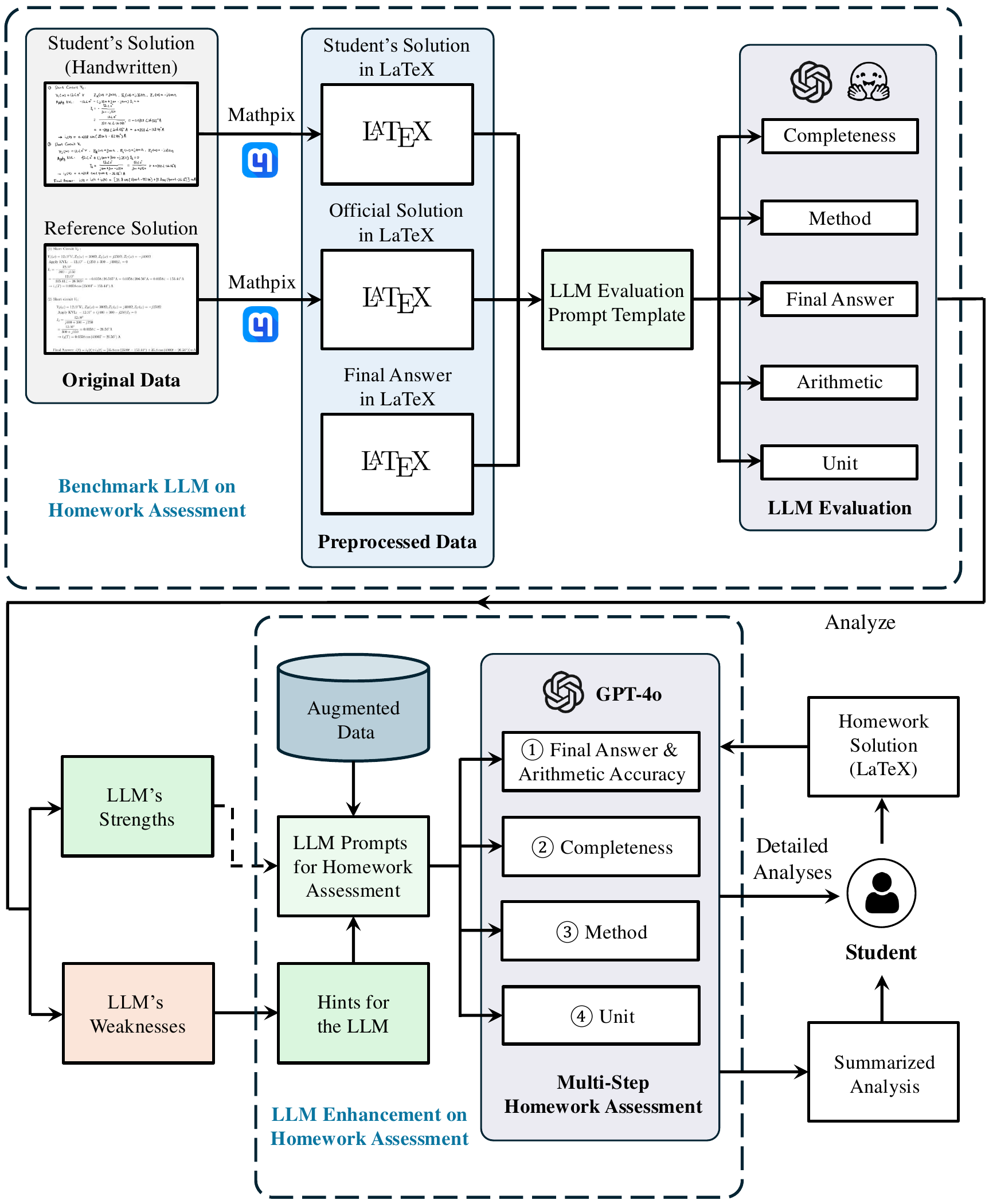}
\caption{Framework for evaluating and enhancing GPT-4o in undergraduate circuit analysis homework assessment. (Note: The evaluation framework shown in the top portion of this figure is adapted from Fig. 1 in \cite{chen2025benchmarking}.)}
\label{fig:method}
\end{figure*}

The remainder of this paper is organized as follows. Section \ref{S3} introduces the preliminaries, with a focus on a benchmarking study of three LLMs—GPT-3.5 Turbo, GPT-4o, and Llama 3 70B—for assessing undergraduate circuit analysis homework. Section \ref{S4} describes the proposed enhancement methods for GPT-4o, including multi-step prompting, context data augmentation, and the use of targeted hints. Section \ref{S5} presents the evaluation results of GPT-4o following the application of these enhancements. Finally, Section \ref{S6} concludes the paper.

\section{Preliminaries: A Benchmarking Study of Large Language Models for Undergraduate Circuit Analysis Homework Assessment}
\label{S3}
In this section, we summarize the preliminary work of this paper: our benchmarking study \cite{chen2025benchmarking} evaluating three LLMs—GPT-3.5 Turbo, GPT-4o, and Llama 3 70B—in the context of providing homework assessments for an undergraduate circuit analysis course. The data collection procedures for both the benchmarking study in \cite{chen2025benchmarking} and the LLM enhancement presented in this paper are detailed in Section \ref{S31}. The metrics used for evaluating the LLMs are introduced in Section \ref{S32}, and the evaluation results are summarized in Section \ref{S33}.

\subsection{Dataset Collections}
\label{S31}
We curated a dataset from an undergraduate circuit analysis course offered by the School of Electrical and Computer Engineering at a large research-intensive university in the Southeastern United States. The dataset comprises 283 handwritten student solutions from 37 students, covering 119 textbook problems across six major topics: i) electric circuit variables and elements, ii) analysis of resistive circuits, iii) the operational amplifier, iv) the complete response of circuits with energy storage elements, v) sinusoidal steady-state analysis, and vi) frequency response. These topics comprehensively cover the contents of an undergraduate-level circuit analysis course. The detailed descriptions of the dataset can be found in \cite{chen2025benchmarking}.

Due to the limitations of current LLMs and vision-language models in interpreting image-based content such as handwritten equations and circuit diagrams, we converted both reference and student solutions into LaTeX using Mathpix\footnote{See https://mathpix.com.}, followed by manual proofreading to ensure accuracy. Circuit diagrams and problem statements were excluded from the LLM evaluation process. However, brief textual descriptions of circuit diagrams will be incorporated during the LLM enhancement stage, as illustrated in Section \ref{S42}. To facilitate evaluation, we also appended concise summaries of each problem’s final answer to help the LLMs assess correctness. As shown in Section \ref{S33}, this approach proved effective for GPT-4o in delivering reliable homework assessments.

\subsection{LLM Evaluation Metrics}
\label{S32}
The evaluated LLMs assess the following aspects of student solutions by comparing them with the corresponding reference solutions:

\begin{itemize}
\item [i)] \textit{Completeness:} This metric assesses whether the student fully addressed all parts of the problem.

\item [ii)] \textit{Method:} This metric evaluates the correctness of the approach or method used, independent of arithmetic errors or typos.

\item [iii)] \textit{Final Answer:} This metric checks the accuracy of the final numerical or symbolic answers provided.

\item [iv)] \textit{Arithmetic:} This metric identifies any arithmetic errors in the solution. When an error is present, LLMs are prompted to specify its nature.

\item [v)] \textit{Unit:} This metric reviews whether appropriate units are used for all variables.
\end{itemize}

The data collection and LLM evaluation pipelines are illustrated in the top portion of Fig. \ref{fig:method}, which is adapted from Fig. 1 in \cite{chen2025benchmarking}.

\subsection{LLM Evaluation Results}
\label{S33}
This section presents the evaluation results of three LLMs—GPT-3.5 Turbo, GPT-4o, and Llama 3 70B—on the dataset introduced in Section \ref{S31}. These models serve as representatives of a basic baseline model, a state-of-the-art closed-source model, and a state-of-the-art open-source model, respectively. To ensure consistency and comparability, all models were evaluated using a standardized prompt template, and their outputs were manually assessed based on predefined grading criteria.

The results show that GPT-4o and Llama 3 70B consistently outperform GPT-3.5 Turbo across all evaluation metrics. GPT-4o, in particular, achieves an average accuracy exceeding 80\%, demonstrating strong capability in grading circuit analysis homework. It also shows consistent performance, maintaining at least 68.97\% accuracy across all topics and metrics. In comparison, Llama 3 70B’s performance is less stable, dropping below 60\% in the unit metric for several advanced topics. Interestingly, although Llama 3 70B slightly outperforms GPT-4o on the completeness and method metrics, a closer examination reveals that this advantage is due in part to its more lenient scoring behavior—it tends to give positive evaluations more readily. GPT-4o applies stricter grading criteria, which leads to a higher false negative rate, especially on metrics where most students perform well. On the other hand, GPT-4o demonstrates superior performance on the unit metric, an area where students frequently make mistakes and where Llama 3 70B’s leniency results in more false positives.

While GPT-4o demonstrates strong overall performance, it is important to examine the types of errors it commonly makes. TABLE \ref{TabGPT4o} categorizes these errors and provides representative examples. In Section \ref{S4}, we introduce enhancement strategies aimed at addressing the identified limitations, with the improved responses shown in TABLE \ref{TabGPT4oEnhanced}.

\begin{table*}
\centering
\caption{Test results of different LLMs across various topics in circuit analysis}
\begin{tabular}{ccccccccc}
\hline
\multirow{2}{*}{LLM} & \multirow{2}{*}{Topic} & \multirow{2}{*}{\makecell{Number \\of Data}} & \multicolumn{6}{c}{Percentages of correct LLM responses in terms of different metrics} \\
\cline{4-9}
& & & Completeness & Method & Final Answer & Arithmetic & Unit & Average\\
\hline
\multirow{7}{*}{\makecell{GPT-3.5 Turbo}} & Var. \& Ele. & 40 & 55.00\% & 87.50\% & 47.50\% & 47.50\% & 72.50\% & 62.00\% \\ 
& Resist. Cir. & 63 & 42.86\% & 73.02\% & 42.86\% & 23.81\% & 55.56\% & 47.62\%\\
& Op. Amp. & 28 & 50.00\% & 78.57\% & 46.43\% & 28.57\% & 53.57\% & 51.43\%\\
& Com. Resp. & 95 & 50.53\% & 69.47\% & 44.21\% & 35.79\% & 49.47\% & 49.89\%\\
& Sinusoidal & 29 & 44.83\% & 86.21\% & 62.07\% & 58.62\% & 58.62\% & 62.07\%\\
& Freq. Resp. & 28 & 42.66\% & 64.29\% & 53.57\% & 25.00\% & 21.43\% & 41.39\%\\\cline{2-9}
& Average & --- & 48.04\% & 74.91\% & 47.35\% & 35.34\% & 52.65\% & 51.66\%\\
\hline
\multirow{7}{*}{\makecell{GPT-4o}} & Var. \& Ele. & 40 & 95.00\% & \textbf{100.00\%} & \textbf{90.00\%} & \textbf{95.00\%} & \textbf{100.00\%} & \textbf{96.00\%}\\ 
& Resist. Cir. & 63 & \textbf{95.24\%} & 92.06\% & 80.95\% & 80.95\% & \textbf{98.41\%} & 89.52\%\\
& Op. Amp. & 28 & 78.57\% & \textbf{96.43\%} & 89.29\% & 82.14\% & \textbf{92.86\%} & 87.86\%\\
& Com. Resp. & 95 & 89.47\% & 96.84\% & \textbf{80.00\%} & \textbf{80.00\%} & \textbf{86.32\%} & \textbf{86.53\%}\\
& Sinusoidal & 29 & 68.97\% & 93.10\% & \textbf{82.76\%} & \textbf{75.86\%} & \textbf{68.97\%} & \textbf{77.93\%}\\
& Freq. Resp. & 28 & \textbf{96.43\%} & 96.43\% & \textbf{92.86\%} & \textbf{92.86\%} & \textbf{92.86\%} & \textbf{94.29\%} \\\cline{2-9}
& Average & --- & 89.05\% & 95.75\% & \textbf{84.10\%} & \textbf{83.39\%} & \textbf{90.46\%} & \textbf{88.55\%}\\
\hline
\multirow{7}{*}{\makecell{Llama 3 70B}} & Var. \& Ele. & 40 & \textbf{97.50\%} & \textbf{100.00\%} & 82.50\% & 75.00\% & 85.00\% & 88.00\%\\ 
& Resist. Cir. & 63 & \textbf{95.24\%} & \textbf{98.41\%} & \textbf{90.48\%} & \textbf{82.54\%} & 92.00\% & \textbf{91.73\%}\\
& Op. Amp. & 28 & \textbf{92.86\%} & 92.86\% & \textbf{92.86\%} & \textbf{89.29\%} & 89.29\% & \textbf{91.43\%}\\
& Com. Resp. & 95 & \textbf{91.58\%} & \textbf{97.89\%} & 78.95\% & 73.68\% & 58.95\% & 80.21\%\\
& Sinusoidal & 29 & \textbf{82.76\%} & \textbf{100.00\%} & \textbf{82.76\%} & 65.52\% & 55.17\% & 77.24\%\\
& Freq. Resp. & 28 & \textbf{96.43\%} & \textbf{100.00\%} & 82.14\% & 89.29\% & 53.57\% & 84.29\%\\\cline{2-9}
& Average & --- & \textbf{92.93\%} & \textbf{98.23\%} & \textbf{84.10\%} & 78.09\% & 72.07\% & 85.08\%\\
\hline
\multicolumn{9}{l}{\makecell[{{p{15.0cm}}}]{* \ding{172} Var. \& Ele. = Electric Circuit Variables and Elements; \ding{173} Resist. Cir. = Analysis of Resistive Circuits; \ding{174} Op. Amp. = The Operational Amplifier; \ding{175} Com. Resp. = The Complete Response of Circuits With Energy Storage Elements; \ding{176} Sinusoidal = Sinusoidal Steady-State Analysis; \ding{177} Freq. Resp. = Frequency Response.\\
** The \textbf{bold} entries indicate the best-performing LLM(s) for each evaluation metric based on the number of correct responses.\\
*** This table is reproduced from Table 3 in \cite{chen2025benchmarking}.}}
\end{tabular}
\label{Tab0}
\end{table*}

\begin{table*}[!t]
\centering
\caption{Notable examples of GPT-4o's inferior response} \label{TabGPT4o}
\begin{tabular}{ccc}
\hline
Type & Example & Note \\
\hline
\multirow{4}{*}{\makecell{Unit Conversion}} & \makecell[{{p{9.25cm}}}]{The correct cost should be \$0.264, but your final answer is 26.4 cents (i.e., \$0.264). It seems you made an error when interpreting units (hours should have been used instead of seconds).} & \makecell[{{p{5.75cm}}}]{GPT-4o fails to identify the equivalence between 26.4 cents and \$0.264.} \\ \cline{2-3}
& \makecell[{{p{9.25cm}}}]{The units for power are inconsistent and incorrect due to the sign convention error. Additionally, the answer is expected in milliwatts (mW), but the unit was given in watts (W) which adds to the confusion.} & \makecell[{{p{5.75cm}}}]{GPT-4o is overly confident that the answer is expected in milliwatts. In fact, the answer given in watts is also acceptable.} \\ \hline
\multirow{6}{*}{\makecell{False\\Incompleteness}} & \makecell[{{p{9.25cm}}}]{The student's solution is incomplete. While they calculated the voltage across the resistors and the power absorbed by each resistor, they did not address the direction of current and adherence to the passive sign convention.} & \makecell[{{p{5.75cm}}}]{The student's solution should be considered complete since they calculated the voltage across the resistors and the power absorbed by each resistor.} \\\cline{2-3}
& \makecell[{{p{9.25cm}}}]{The student's solution is incomplete. While it attempts to find the current ($i$) and voltage ($v_o$), it lacks a detailed step-by-step explanation which includes the intermediate steps and proper application of Kirchhoff's Voltage Law (KVL) and Kirchhoff's Current Law (KCL). The discussion concerning $i_2$ and intermediate voltages like $v_a$ is missing.} & \makecell[{{p{5.75cm}}}]{The student's solution is actually complete but misses some step-by-step explanations.} \\
\hline
\multirow{4.35}{*}{\makecell{Number Format}} & \makecell[{{p{9.25cm}}}]{Unfortunately, your final answer for \( R_t \) is incorrect. The official answer is \( R_t = -\frac{8}{3} \, \Omega \), but you calculated it as \( -\frac{16}{6} \, \Omega \).} & \makecell[{{p{5.75cm}}}]{GPT-4o fails to recognize the equivalence between $-\frac{8}{3}~\Omega$ and $-\frac{16}{6}~\Omega$.} \\
\cline{2-3}
& \makecell[{{p{9.25cm}}}]{The student's final answer is: $V(t)=\sqrt{74} \cos \left(8 t-9.462^{\circ}\right)$\\
While the correct answer should be: $v(t)=8.602 \cos \left(8 t-9.462^{\circ}\right) ~\text{V}$\\
It can be observed that the student's amplitude calculation is incorrect.} & \makecell[{{p{5.75cm}}}]{Since \(\sqrt{74} \approx 8.602\), the student's final answer matches the official solution up to a rounding error.} \\
\hline
\makecell{Alternative\\Method} & \makecell[{{p{9.25cm}}}]{The student applies KCL to find \(R_t\). However, the method and steps are somewhat unclear and differ from the standard approach usually taught, which directly utilizes the short-circuit current \(i_{sc}\) and open-circuit voltage \(v_{oc}\). This might be confusing.} & \makecell[{{p{5.75cm}}}]{The student's solution uses an alternative approach to find $R_{t}$. Although the method used is different from the approach in the official solution, it is also correct.} \\
\hline
\makecell{Different\\Notations} & \makecell[{{p{9.25cm}}}]{There are arithmetic errors in your calculations for \( i_1 \) and \( i_2 \). \\
-- Correct calculation:\\$i_1=\frac{8}{16+8}(-6)=-2 \text{ A}$, $i_2=\frac{8}{8+8}(-6)=-3 \text{ A}$\\
-- Your incorrect calculation:\\$i_1=\frac{1 / 8}{\frac{1}{8}+\frac{1}{8}} \cdot 6 \text{ A}=3 \text{ A}$, $i_2=\frac{1 / 16}{ \frac{1}{16}+\frac{1}{8}} \cdot 6 \text{ A}=2 \text{ A}
$} & \makecell[{{p{5.75cm}}}]{GPT-4o fails to recognize that the student's solution uses a different notation system from the official solution. The currents \( i_{1} \) and \( i_{2} \) in the student's solution represent \( i_{2} \) and \( i_{1} \), respectively, in the official solution. GPT-4o cannot recognize the exchanged variable notations in this example.} \\
\hline
\multirow{18}{*}{\makecell{Arithmetic}} & \makecell[{{p{9.25cm}}}]{There are errors in the arithmetic and the process leading to the calculation, primarily in their steps involving the arithmetic for current division:\\
-- For the 9 A source: $i_{20A} = 9\text{~A} \cdot \frac{1/35\Omega}{1/10\Omega + 1/35\Omega} \neq 2\text{~A}$\\
Correct method: $i_{20A} = 9\text{~A} \cdot \frac{10\Omega}{10\Omega + 35\Omega} = 9\text{~A} \cdot \frac{10}{45} = 9\text{~A} \cdot \frac{2}{9} = 2\text{~A}$\\
-- For the 6 A source: $i_{20B} = 6\text{~A} \cdot \frac{1/30\Omega}{1/30\Omega + 1/15\Omega} \neq 2\text{~A}$\\
Correct method: $i_{20B} = 6\text{~A} \cdot \frac{15\Omega}{15\Omega + 30\Omega} = 6\text{~A} \cdot \frac{15}{45} = 6\text{~A} \cdot \frac{1}{3} = 2\text{~A}$} & \makecell[{{p{5.75cm}}}]{Although GPT-4o sometimes performs detailed mathematical calculations, the results are not necessarily correct. In this example, we should have $9\text{~A} \cdot \frac{1/35\Omega}{1/10\Omega + 1/35\Omega} = 2\text{~A}$ and $6\text{~A} \cdot \frac{1/30\Omega}{1/30\Omega + 1/15\Omega} = 2\text{~A}$.} \\
\cline{2-3}
& \makecell[{{p{9.25cm}}}]{There are no significant arithmetic errors that affect the final result. However, there is a sign issue to note:\\
-- In the equation for \( \text{KCL@} V_b \), you have correctly identified the expression \( \frac{-V_3}{6000}-\frac{V_0}{6000}=0 \), but then you incorrectly simplified it to \( -4 - V_0 = 0 \). The correct simplification should be \( -\frac{4}{6000} - \frac{V_0}{6000} = 0 \) which equates to \( -4 - V_0 = 0 \), so this appears to be more of a transcription error than an arithmetic error.} & \makecell[{{p{5.75cm}}}]{The response of GPT-4o is redundant since it is completely fine to derive \( -4 - V_0 = 0 \) directly from \( \frac{-V_3}{6000} - \frac{V_0}{6000} = 0 \), given \( V_{3} = 4 \) V.} \\
\cline{2-3}
& \makecell[{{p{9.25cm}}}]{The student's final expression for \( V_0 \) is: \\$V_0 = \frac{R_1(V_1 - V_2) - R_3(V_2 - V_1)}{R_2} + V_1 - V_2$\\
Simplifying this, \\
$V_0 = \frac{R_1(V_1 - V_2) + R_3(V_1 - V_2)}{R_2} + V_1 - V_2$, 
$V_0 = \frac{(R_1 + R_3)(V_1 - V_2)}{R_2} + V_1 - V_2$\\
This does not match the official solution: $v_0 = \frac{R_1 + R_2 + R_3}{R_2}(v_1 - v_2)$\\
Thus, the student's final answer is incorrect. The steps might have errors that cause the final discrepancy.} & \makecell[{{p{5.75cm}}}]{GPT-4o has some level of calculation ability. However, it does not fully simplify the original equation for \( V_{0} \) and thus made an incorrect conclusion that the student's final expression does not match the official solution.} \\
\cline{2-3}
& \makecell[{{p{9.25cm}}}]{The student's final answer $i(t) = -3 \sin(2t + 30^\circ)$ is not in the correct form. The official solution is given by $i(t) = 3 \cos(2t + 120^\circ)$. There is an error in converting the trigonometric function from sine to cosine form.} & \makecell[{{p{5.75cm}}}]{GPT-4o does not identify the equivalence between \(-3 \sin(2t + 30^\circ)\) and \(3 \cos(2t + 120^\circ)\).} \\
\hline
\makecell{Rounding Errors} & \makecell[{{p{9.25cm}}}]{The student's final answer \( 14.69 \angle 5.55^{\circ} \) is close but not correct. The correct answer is \( 14.67 \angle 5.6^{\circ} \).} & \makecell[{{p{5.75cm}}}]{GPT-4o does not recognize the value discrepancies in the student's final answer as rounding errors.} \\
\hline
\makecell{Term Order\\in Equations} & \makecell[{{p{9.25cm}}}]{The final equation \( 1 \times 10^8 v_s = \frac{d^2 v}{dt^2} + 3000 \frac{dv}{dt} + 1.02 \times 10^8 v_a \) the student reaches is incorrect primarily because of coefficient and term sign errors throughout the derivation. This divergence from the correct answer \( 1 \times 10^8 v_{\mathrm{s}} = 1.02 \times 10^8 v + 3000 \frac{\mathrm{d}v}{\mathrm{d}t} + \frac{\mathrm{d}^2 v}{\mathrm{d}t^2} \) indicates a need to revisit algebraic and differential handling within this setup.} & \makecell[{{p{5.75cm}}}]{GPT-4o fails to recognize the equivalence between two differential equations in which the terms are arranged in different orders.} \\
\hline
\makecell{Hallucinations} & \makecell[{{p{9.25cm}}}]{-- Consistently use proper notation for complex numbers and phasors.\\
-- Provide final results in correct units — Volts for $\mathbf{V}_{\mathrm{oc}}$ and Ohms for $\mathbf{Z}_{\mathrm{t}}$. } & \makecell[{{p{5.75cm}}}]{The student, in effect, did not consistently use units for all variables.} \\
\hline
\multicolumn{3}{l}{\makecell[{{p{17cm}}}]{* This table is reproduced from Table 10 in \cite{chen2025benchmarking}.}}
\end{tabular}
\end{table*}

\section{Enhancing GPT-4o for Homework Assessment}
\label{S4}
Section \ref{S3} shows that GPT-4o achieves relatively strong overall performance. However, as summarized in TABLE \ref{TabGPT4o}, GPT-4o still exhibits several recurring errors. Reliability is a critical factor in generating homework assessments and feedback, as incorrect responses can mislead students and negatively affect their learning outcomes—especially for those who are already struggling \cite{rohde2024predictors, sun2025data}. In this section, we aim to enhance GPT-4o’s performance in circuit analysis homework assessment by employing several strategies, including multi-step prompting, context data augmentation, and the incorporation of targeted hints.

\begin{figure*}
\centering
\includegraphics[width=\textwidth]{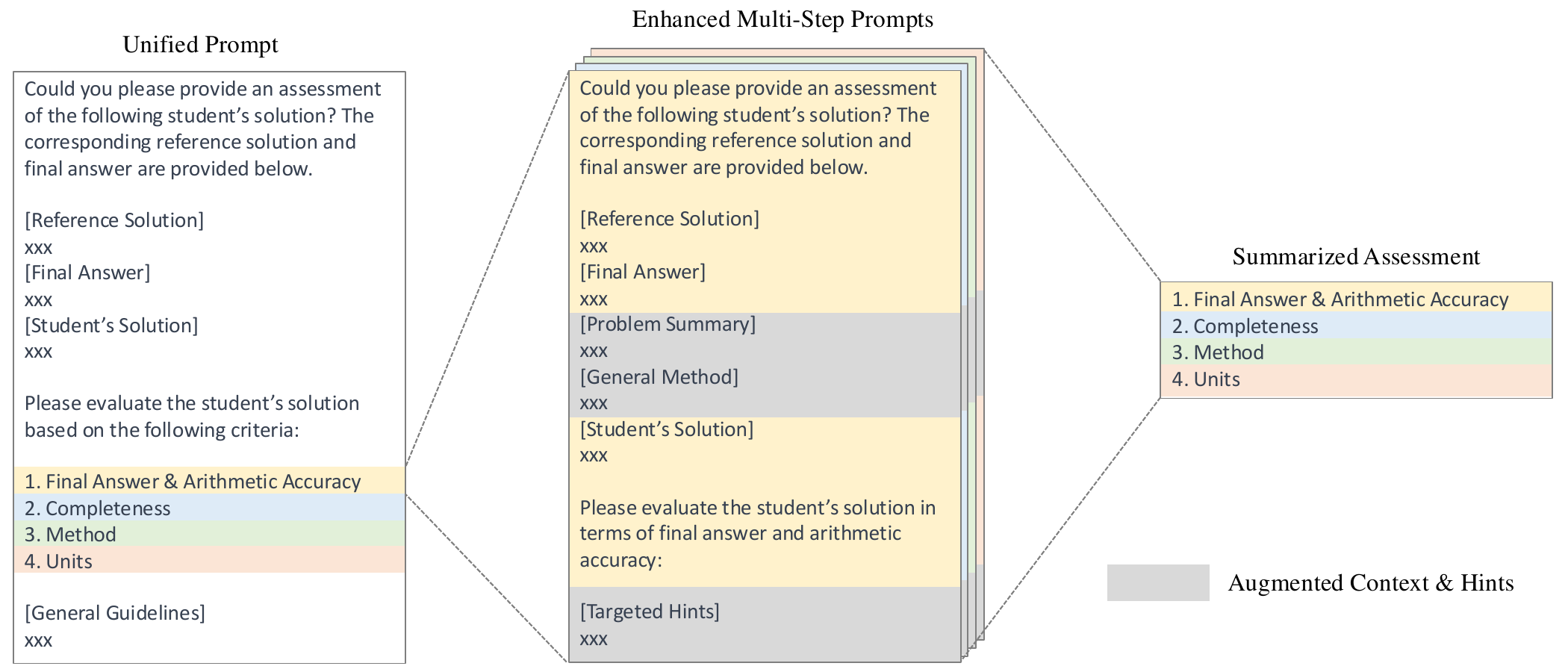}
\caption{Enhancement of GPT-4o in undergraduate circuit analysis homework assessment. (Note: ``xxx" represents the contents of the corresponding items.)}
\label{fig:enhancement}
\end{figure*}

\begin{table*}[!h]
\centering
\caption{Appended hints for different prompting steps}
\label{Tab:hints}
\begin{tabular}{ccl}
\hline
Evaluation Aspect & Error Type & \makecell{Hint} \\
\hline
\multirow{18}{*}{\makecell{Final answer \&\\ arithmetic accuracy}} & \multirow{3}{*}{Arithmetic} & You need to conduct a careful inspection when comparing the final answers. \\ \cline{3-3}
& & \makecell[{{p{12.5cm}}}]{The numbers and their signs must be correct, and the errors of them cannot be regarded as typos. If there are any sign errors, you need to make to point them out.} \\ \cline{2-3}
& Rounding Errors & \makecell[{{p{12.5cm}}}]{Rounding errors in the calculation process should not be treated as calculation errors.}\\ \cline{2-3}
& \multirow{10.5}{*}{Number Format} & \makecell[{{p{12.5cm}}}]{You should recognize equivalencies between different number formats, such as scientific and standard notation. For instance, i) $3333.33$ is equivalent to $3.33 \times 10^{{3}}$; ii) $\sqrt{{74}}$ is equivalent to $8.602$.}\\ \cline{3-3}
& & \makecell[{{p{12.5cm}}}]{You should recognize the equivalency between fractions and decimals. For example, $\frac{{1}}{{3}}$ is equivalent to $0.333$, accounting for rounding error.}\\ \cline{3-3}
& & \makecell[{{p{12.5cm}}}]{You should recognize the equivalency between fractions after reduction. For example, $\frac{{10}}{{6}}$ is equivalent to $\frac{{5}}{{3}}$.}\\ \cline{3-3}
& & \makecell[{{p{12.5cm}}}]{You should recognize the equivalency of final answers under trigonometric identities. For example, $\sin(2t + 30^{{\circ}}) = -\cos(2t + 120^{{\circ}})$. In order to identify the equivalency like this, you should draw detailed and step-by-step methematical deductions.}\\ \cline{3-3}
& & \makecell[{{p{12.5cm}}}]{You should recognize the equivalency between degrees and radians. For example, $\frac{{\pi}}{{3}}$ is equivalent to $60^{{\circ}}$.}\\ \cline{3-3}
& & \makecell[{{p{12.5cm}}}]{You should recognize the equivalency of terms under the commutative property.}\\ \cline{3-3}
& & \makecell[{{p{12.5cm}}}]{You should recognize the equivalency between notations involving imaginary units. For example, $-j$ is equivalent to $\frac{{1}}{{j}}$.}\\ \cline{2-3}
& \makecell{Term Order\\in Equations} & \makecell[{{p{12.5cm}}}]{When comparing differential equations, first ensure the terms are in the same order by making conversions. Then, compare their coefficients.}\\ 
\hline
\multirow{1}{*}{Completeness} & \makecell{False\\Incompleteness} & \makecell[{{p{12.5cm}}}]{The student's solution should be considered as complete as long as it answers the question in the problem, no matter whether the answer is correct or not.} \\
\hline 
\multirow{3.25}{*}{Method} & General & \makecell[{{p{12.5cm}}}]{You do not need to check the details. The used method can be considered as correct as long as it is generally correct.}\\ \cline{2-3}
& \makecell{Alternative\\Method} & \makecell[{{p{12.5cm}}}]{The student may use an alternative method. If the final answer is correct, assume the method used was valid. If the final answer is incorrect, carefully determine whether the student's approach was valid.} \\
\hline
\multirow{7}{*}{Units} & \multirow{2}{*}{General} & \makecell[{{p{12.5cm}}}]{In calculations involving notations, units may be implicit. Missing units in such cases should not be considered errors, especially in topics like frequency responses.}\\ \cline{3-3}
& & \makecell[{{p{12.5cm}}}]{The units can be implicitly used in the intermediate steps.}\\ \cline{2-3}
& \multirow{3.25}{*}{Unit Conversion} & \makecell[{{p{12.5cm}}}]{If the student's answer uses a different unit from the reference, convert them to the same unit for comparison. For example, 26.4 cents equals 0.264 dollars. If the answer is correct after conversion, the unit should also be considered correct.}\\ \cline{3-3}
& & \makecell[{{p{12.5cm}}}]{You should interpret the units with flexibility. For example, ``second" is a same unit with ``s" when they are referring to time.}\\ 
\hline
\end{tabular}
\end{table*}

\subsection{Multi-Step Prompting for Different Aspects}
\label{S41}
The GPT-4o evaluations in Section \ref{S3} employ a unified prompt template to assess students' homework solutions across five metrics. However, a key drawback of this approach is that GPT-4o must consider all aspects of the assessment simultaneously, which can lead to a loss of focus and an increased likelihood of hallucinations \cite{liu2024lost}.

We adopt a multi-step prompting strategy to guide GPT-4o in generating homework assessments from different perspectives. Each step includes the reference solution, the student’s solution, and other problem-specific data (see Section \ref{S42}) in the prompt, and focuses on evaluating a specific aspect of the student’s work. Since a correct final answer often implies the absence of arithmetic errors, the evaluation of arithmetic calculations and final answers is combined into a single step. As a result, the evaluated metrics across all steps include: i) final answer and arithmetic accuracy, ii) completeness, iii) method, and iv) units. By isolating each evaluation focus, the multi-step prompting approach helps mitigate GPT-4o’s tendency to hallucinate. Additionally, we generate a summary of GPT-4o’s assessments across all aspects, enabling students to review a concise overall evaluation without having to read each detailed response individually. The LLM enhancement approach using multi-step prompting is illustrated in Fig. \ref{fig:enhancement}.

\subsection{Context Data Augmentation}
\label{S42}
For the GPT-4o evaluations in Section \ref{S3}, the problem-specific context includes both the reference solution and the final answer. GPT-4o generates homework assessments by comparing the student’s solution to the corresponding reference solution, while the final answer provides clarity on which variables the problem requires students to solve.

Since much of the problem context is typically embedded within circuit diagrams, we explicitly include summaries of the problem descriptions. These summaries provide brief descriptions of the circuit diagrams and identify the variables to be solved, thereby offering additional problem-specific context. Furthermore, we include the general problem-solving method—such as mesh current analysis or node voltage analysis—in the prompt to clearly inform GPT-4o of the intended analytical approach. This enriched context enhances GPT-4o’s evaluation performance. Fig. \ref{fig:enhancement} illustrates how the augmented context is incorporated into the multi-step prompts.

\begin{table*}[!h]
\centering
\caption{GPT-4o's enhanced responses to the examples in TABLE \ref{TabGPT4o}} \label{TabGPT4oEnhanced}
\begin{tabular}{ccc}
\hline
Type & Original snippet from GPT-4o's evaluations & Enhanced version of GPT-4o's response \\
\hline
\multirow{5}{*}{\makecell{Unit Conversion}} & \makecell[{{p{7.5cm}}}]{\vspace{-0.2cm} {\color{red}\ding{55}} The correct cost should be \$0.264, but your final answer is 26.4 cents (i.e., \$0.264). It seems you made an error when interpreting units (hours should have been used instead of seconds).} & \makecell[{{p{7.5cm}}}]{\vspace{-0.2cm}{\color{green}\ding{52}}{\color{red}\ding{55}} You calculated the energy consumption correctly as 2.2 kWh, but when you calculated the cost, you used 12 cents per kWh instead of the given rate.} \\ \cline{2-3}
& \makecell[{{p{7.5cm}}}]{\vspace{-0.2cm} {\color{red}\ding{55}} The units for power are inconsistent and incorrect due to the sign convention error. Additionally, the answer is expected in milliwatts (mW), but the unit was given in watts (W) which adds to the confusion.} & \makecell[{{p{7.5cm}}}]{\vspace{-0.2cm}{\color{green}\ding{52}} The units are consistent here, but the values used are incorrect based on the problem's given data.} \\ \hline
\multirow{7}{*}{\makecell{False\\Incompleteness}} & \makecell[{{p{7.5cm}}}]{\vspace{-0.2cm} {\color{red}\ding{55}} The student's solution is incomplete. While they calculated the voltage across the resistors and the power absorbed by each resistor, they did not address the direction of current and adherence to the passive sign convention.} & \makecell[{{p{7.5cm}}}]{\vspace{-0.175cm}{\color{green}\ding{52}} Yes, your solution is complete because you addressed all parts of the problem by calculating the voltages across the resistors and the power absorbed by each.} \\\cline{2-3}
& \makecell[{{p{7.5cm}}}]{\vspace{-0.2cm} {\color{red}\ding{55}} The student's solution is incomplete. While it attempts to find the current ($i$) and voltage ($v_o$), it lacks a detailed step-by-step explanation which includes the intermediate steps and proper application of Kirchhoff's Voltage Law (KVL) and Kirchhoff's Current Law (KCL). The discussion concerning $i_2$ and intermediate voltages like $v_a$ is missing.} & \makecell[{{p{7.5cm}}}]{\vspace{-0.15cm}{\color{green}\ding{52}} Yes, your answer is complete! You've addressed both the voltage and current as required by the problem.} \\
\hline
\multirow{5.25}{*}{\makecell{Number Format}} & \makecell[{{p{7.5cm}}}]{\vspace{-0.2cm} {\color{red}\ding{55}} Unfortunately, your final answer for \( R_t \) is incorrect. The official answer is \( R_t = -\frac{8}{3} \, \Omega \), but you calculated it as \( -\frac{16}{6} \, \Omega \).} & \makecell[{{p{7.5cm}}}]{\vspace{-0.2cm}{\color{green}\ding{52}} When you calculated $R_{Th}$, you ended up with $-\frac{16}{6}~\Omega$. Double-check your division there, as it should simplify to a different fraction.} \\
\cline{2-3}
& \makecell[{{p{7.5cm}}}]{\vspace{-0.2cm} {\color{red}\ding{55}} The student's final answer is: $V(t)=\sqrt{74} \cos \left(8 t-9.462^{\circ}\right)$\\
While the correct answer should be: $v(t)=8.602 \cos (8 t-9.462^{\circ}) ~\text{V}$. It can be observed that the student's amplitude calculation is incorrect.} & \makecell[{{p{7.5cm}}}]{\vspace{-0.175cm}{\color{green}\ding{52}} Your final expression for $v(t)$ is $\sqrt{74} \cos \left(8 t - 9.462^{\circ}\right)$. This is equivalent to the reference solution, as $\sqrt{74}$ is approximately $8.602$. So, your magnitude and phase are correct.} \\
\hline
\makecell{Alternative\\Method} & \makecell[{{p{7.5cm}}}]{\vspace{-0.2cm} {\color{red}\ding{55}} The student applies KCL to find \(R_t\). However, the method and steps are somewhat unclear and differ from the standard approach usually taught, which directly utilizes the short-circuit current \(i_{sc}\) and open-circuit voltage \(v_{oc}\). This might be confusing.} & \makecell[{{p{7.5cm}}}]{\vspace{-0.175cm}{\color{green}\ding{52}} The method you've used to find the open-circuit voltage and the Th'evenin resistance seems generally correct. You've applied KCL and KVL effectively to solve for the necessary values.} \\
\hline
\multirow{21}{*}{\makecell{Arithmetic}} & \makecell[{{p{7.5cm}}}]{\vspace{-0.2cm} {\color{red}\ding{55}} There are errors in the arithmetic and the process leading to the calculation, primarily in their steps involving the arithmetic for current division:\\
-- For the 9 A source: $i_{20A} = 9\text{~A} \cdot \frac{1/35\Omega}{1/10\Omega + 1/35\Omega} \neq 2\text{~A}$. Correct method: $i_{20A} = 9\text{~A} \cdot \frac{10\Omega}{10\Omega + 35\Omega} = 9\text{~A} \cdot \frac{10}{45} = 9\text{~A} \cdot \frac{2}{9} = 2\text{~A}$\\
-- For the 6 A source: $i_{20B} = 6\text{~A} \cdot \frac{1/30\Omega}{1/30\Omega + 1/15\Omega} \neq 2\text{~A}$. Correct method: $i_{20B} = 6\text{~A} \cdot \frac{15\Omega}{15\Omega + 30\Omega} = 6\text{~A} \cdot \frac{15}{45} = 6\text{~A} \cdot \frac{1}{3} = 2\text{~A}$} & \makecell[{{p{7.5cm}}}]{\vspace{-0.15cm}{\color{green}\ding{52}} There are no arithmetic errors in the student's solution. They correctly applied superposition and current division to find the voltage across the 20 $\Omega$ resistor for each current source separately, and then summed the results accurately.} \\
\cline{2-3}
& \makecell[{{p{7.5cm}}}]{\vspace{-0.2cm} {\color{red}\ding{55}} There are no significant arithmetic errors that affect the final result. However, there is a sign issue to note:\\
-- In the equation for \( \text{KCL@} V_b \), you have correctly identified the expression \( \frac{-V_3}{6000}-\frac{V_0}{6000}=0 \), but then you incorrectly simplified it to \( -4 - V_0 = 0 \). The correct simplification should be \( -\frac{4}{6000} - \frac{V_0}{6000} = 0 \) which equates to \( -4 - V_0 = 0 \), so this appears to be more of a transcription error than an arithmetic error.} & \makecell[{{p{7.5cm}}}]{\vspace{-0.15cm}{\color{green}\ding{52}} 1. For the final answers, you've correctly determined that $v_{\text{o}} = -4 \text{ V}$ and $i_{\text{o}} = 1.33 \text{ mA}$. The signs and values match perfectly with what is needed in the given problem. \\ 2. There aren't any arithmetic errors in your calculations. You've applied KCL correctly at each node and solved for the voltages and currents accurately.} \\
\cline{2-3}
& \makecell[{{p{7.5cm}}}]{\vspace{-0.2cm} {\color{red}\ding{55}} The student's final expression for \( V_0 \) is: \\$V_0 = \frac{R_1(V_1 - V_2) - R_3(V_2 - V_1)}{R_2} + V_1 - V_2$. \\
Simplifying this, $V_0 = \frac{R_1(V_1 - V_2) + R_3(V_1 - V_2)}{R_2} + V_1 - V_2$, $V_0 = \frac{(R_1 + R_3)(V_1 - V_2)}{R_2} + V_1 - V_2$. \\ This does not match the official solution: $v_0 = \frac{R_1 + R_2 + R_3}{R_2}(v_1 - v_2)$. Thus, the student's final answer is incorrect. The steps might have errors that cause the final discrepancy.} & \makecell[{{p{7.5cm}}}]{\vspace{-0.2cm} {\color{red}\ding{55}} It looks like you've made a good effort in solving the problem, but there seems to be a mistake in the final expression for $v_0$. The reference solution expresses $v_0$ as a function of $v_1$ and $v_2$ with a specific combination of resistors. Your final expression doesn't match this form exactly, particularly in how the resistors are combined.} \\
\cline{2-3}
& \makecell[{{p{7.5cm}}}]{\vspace{-0.2cm} {\color{red}\ding{55}} The student's final answer $i(t) = -3 \sin(2t + 30^\circ)$ is not in the correct form. The official solution is given by $i(t) = 3 \cos(2t + 120^\circ)$. There is an error in converting the trigonometric function from sine to cosine form.} & \makecell[{{p{7.5cm}}}]{\vspace{-0.2cm}{\color{green}\ding{52}} You're doing well with your calculations! You applied the method correctly by using the relationship between voltage and current in a capacitor and took the derivative properly. Your expression for $i(t) = -3 \sin(2t + 30^\circ)$ is correct.} \\
\hline
\makecell{Rounding Errors} & \makecell[{{p{7.5cm}}}]{\vspace{-0.2cm} {\color{red}\ding{55}} The student's final answer \( 14.69 \angle 5.55^{\circ} \) is close but not correct. The correct answer is \( 14.67 \angle 5.6^{\circ} \).} & \makecell[{{p{7.5cm}}}]{\vspace{-0.2cm}{\color{green}\ding{52}} Your final answer for $v_{\rm{o}}(t) = 14.69 \cos \left(2500 t + 5.55^{\circ}\right)$ is very close to the reference solution. The magnitude and phase are slightly different, but they are within a reasonable range, likely due to rounding differences.} \\
\hline
\makecell{Term Order\\in Equations} & \makecell[{{p{7.5cm}}}]{\vspace{-0.2cm} {\color{red}\ding{55}} The final equation \( 1 \times 10^8 v_s = \frac{d^2 v}{dt^2} + 3000 \frac{dv}{dt} + 1.02 \times 10^8 v_a \) the student reaches is incorrect primarily because of coefficient and term sign errors throughout the derivation.\\This divergence from the correct answer \( 1 \times 10^8 v_{\mathrm{s}} = 1.02 \times 10^8 v + 3000 \frac{\mathrm{d}v}{\mathrm{d}t} + \frac{\mathrm{d}^2 v}{\mathrm{d}t^2} \) indicates a need to revisit algebraic and differential handling within this setup.} & \makecell[{{p{7.5cm}}}]{\vspace{-0.2cm}{\color{green}\ding{52}} Final Answer: Your final differential equation matches the reference solution in terms of structure and coefficients. The terms are in the correct order, and the coefficients are accurate.\\ Arithmetic Errors: I didn't find any arithmetic errors in your calculations. You applied KCL and KVL correctly and substituted the values accurately.} \\
\hline
\end{tabular}
\end{table*}

\subsection{Incorporation of Targeted Hints}
\label{S43}
TABLE \ref{TabGPT4o} presents examples of GPT-4o’s responses that contain various errors. These errors illustrate the types of mistakes GPT-4o is prone to when provided with general homework assessment guidance. To reduce the frequency of such errors, an effective strategy is to incorporate targeted hints into the prompt, explicitly guiding GPT-4o to avoid them. Using the multi-step prompting method described in Section \ref{S41}, these hints can be integrated into specific steps, thereby avoiding overly lengthy prompts. TABLE \ref{Tab:hints} lists the hints appended to different prompting stages, each designed to address the common errors shown in TABLE \ref{TabGPT4o}. To resolve notation inconsistencies, circuit variable notations can be clearly specified in the problem statements, encouraging students to use a notation system consistent with the reference solutions. Given that LLMs are susceptible to hallucinations, we aim to mitigate this issue in future work through advanced techniques such as model fine-tuning \cite{han2024parameter} and retrieval-augmented generation (RAG) \cite{gao2023retrieval}.

\section{GPT-4o Enhancement Results}
\label{S5}
To demonstrate the enhanced performance of GPT-4o using the strategies described in Section \ref{S4}, we evaluated GPT-4o with the new prompting method and augmented data on the examples shown in TABLE \ref{TabGPT4o}, which include cases where GPT-4o previously produced incorrect responses during evaluations. The improved responses are presented in TABLE \ref{TabGPT4oEnhanced}, where the symbols ``{\color{green}\ding{52}}" and ``{\color{red}\ding{55}}" indicate correct and incorrect responses, respectively. In the first example on unit conversions (marked with ``{\color{green}\ding{52}}{\color{red}\ding{55}}"), GPT-4o does not mention that the student’s final answer is incorrect due to unit issues—which is the desired behavior. However, the response contains an additional error resulting from a general hallucination. In the third arithmetic example, GPT-4o still fails to provide a correct assessment of a student’s solution involving notation calculations, highlighting its limited intrinsic capabilities in this area. Notably, of the 13 examples in which GPT-4o’s original responses contained errors, 11 were correctly addressed using the enhanced methods. In future work, we plan to explore more advanced techniques to further improve the quality of GPT-4o’s responses. 

Additionally, we tested the new prompting strategy using 87 raw student solutions, of which 39 are from the topic of electric circuit variables and elements (i.e., Chapters 1 and 2 in \cite{svoboda2013introduction}) and 48 are from the analysis of resistive circuits (i.e., Chapters 3 and 4 in \cite{svoboda2013introduction}). TABLE \ref{Tab2} presents the numbers and percentages of correct responses from GPT-4o with and without the enhancement strategies. A response is considered correct if GPT-4o accurately evaluates \textit{all} aspects of the student’s solution. As shown in the table, the correct response rate increases from 74.71\% to 97.70\% after applying the enhanced prompting and augmented data, demonstrating the effectiveness of the enhancement methods described in Section \ref{S4}.

\begin{table}[!t]
	\centering
	\caption{Numbers and percentages of correct responses from GPT-4o with and without the enhancement strategies}
	\label{Tab2}
	\begin{tabular}{ccc}
		\hline
		Prompt & Correct response number & Correct response ratio\\
		\hline
		w/o enhancement & 65 & 74.71\% \\
		w/ enhancement & 85 & 97.70\% \\
		\hline
	\end{tabular}
\end{table}

\section{Conclusions}
\label{S6}
This study demonstrates the promising potential of LLMs in supporting automated assessment for undergraduate circuit analysis coursework. Building on GPT-4o’s strong baseline performance, we proposed a structured enhancement framework involving multi-step prompting, context data augmentation, and incorporation of targeted hints. These strategies significantly improve the accuracy and reliability of assessments, especially in entry-level topics, pushing the correct response rate of GPT-4o from 74.71\% to 97.70\%. The findings highlight both the current capabilities and limitations of LLMs in educational settings and offer a pathway toward developing robust, scalable tools for engineering instruction \cite{chen2025wip}. Future work will explore integrating information from circuit diagrams into the framework to reduce reliance on manually prepared data. We also aim to extend these techniques to more advanced topics and other engineering domains, further embedding LLMs into the educational workflow as effective and intelligent teaching assistants.

\section{Acknowledgments}
The authors appreciate the support provided by the School of Electrical and Computer Engineering and the College of Engineering at the institution where the study was conducted. The authors would also like to acknowledge the assistance of ChatGPT in polishing the language of this paper.

\bibliographystyle{IEEEtran}
\bibliography{IEEEfull.bib}

\end{document}